# Distributed Shared Layered Storage Quantum Simulator: A novel quantum simulation system for efficient scaling and cost optimization

Mingyang Yu, Haorui Yang, Donglin Wang, Desheng Kong, Ji Du, Yulong Fu, Wei Wang and Jing Xu, *Member, IEEE*

*Abstract*—Quantum simulators are essential tools for developing and testing quantum algorithms. However, the high-frequency traversal characteristic of quantum simulators represents an unprecedented demand in the history of IT, and existing distributed technologies is unable to meet this requirement, resulting in a single-node bottleneck of quantum simulator. To overcome this limitation, this paper introduces a novel Distributed Shared Layered Storage Quantum Simulator (DSLSQS). By leveraging an innovative distributed architecture in which multiple computational nodes share data storage directly, together with De-TCP/IP networking technology, DSLSQS effectively eliminates East-West data flow in distributed systems. This approach mitigates the bottleneck of distributed quantum simulation clusters and enhances the scalability. Moreover, the system employs layered storage technology, which reduces usage of expensive high-performance memory and substantially lowers simulation costs. Furthermore, this paper systematically analyzes the performance and cost constraints of distributed quantum simulator cluster, identifying distributed networking as the primary performance bottleneck and highlighting that minimizing storage costs is crucial to reducing the total cost. Finally, experimental evaluations with a 27-qubit simulation confirm the successful implementation of layered storage within the quantum simulator. DSLSQS significantly enhances simulation efficiency, yielding a performance improvement of over 350% compared to existing distributed technologies. These results underscore the superior performance and scalability of the proposed architecture in managing complex quantum computing tasks. This paper provides crucial insights for the practical deployment of quantum computing and presents an effective framework for the development of distributed quantum simulation clusters.

## I. Introduction

With the rapid advancement of quantum technologies, quantum computing has garnered significant attention from both academia and industry [1]. Unlike classical computing, quantum computing leverages the fundamental principles of quantum mechanics, such as superposition and entanglement [2], to enable exponential computational speedup. This capability allows for solving problems that are computationally infeasible for classical computers, such as large integer factorization and the simulation of complex quantum systems [3]. However, the development of quantum computers remains constrained by several technical challenges, including qubit decoherence, the complexity of quantum error correction, and limitations in hardware scalability, which hinder the realization of large-scale, practical quantum systems [4]. Quantum simulators facilitate the emulation of quantum computations on classical hardware, enabling researchers to validate quantum algorithms, optimize quantum software, and investigate the fundamental properties of quantum systems [5, 6]. Such research has not only driven theoretical advancements in quantum computing but has also provided essential insights for the design and enhancement of quantum hardware.

A classical quantum simulator is a computational tool designed to emulate the behavior of a quantum computer using conventional classical hardware. This simulator leverages classical computing resources (e.g., CPUs or GPUs) to model qubit states and quantum gate operations [7], thereby facilitating quantum computing experiments and research in classical computational environments [8]. A full-amplitude quantum simulator executes computations by explicitly modeling the complete amplitudes of quantum states, thereby allowing for the simulation of arbitrary quantum circuits. However, this approach necessitates storing and processing the amplitudes of all possible quantum states. With each additional qubit, the required data space doubles, and every quantum gate operation must traverse this exponentially growing state space. While this process remains feasible for small-scale quantum systems, as the number of qubits grows, the volume of data that must be processed within a limited timeframe increases exponentially, resulting in a rapid escalation of storage and data throughput demands. Consequently, the strain on storage and data bandwidth intensifies significantly.

The prevailing architecture for quantum simulators relies on a single server, where all data is stored within the server's memory [9]. However, the capacity of such simulators is fundamentally constrained by the computational resources of a single server, leading to a scalability bottleneck at approximately 34 qubits.

To overcome this challenge, distributed scaling is considered a pivotal technological breakthrough in quantum simulation [10], paving the way for the commercialization of quantum simulators. However, existing distributed technologies fall short of meeting the unique demands of quantum simulators, which require traversing vast quantum state information for each quantum gate. The necessity for high-frequency traversal of extensive data sets places substantial strain on the system's internal East-West data traffic bandwidth. Existing distributed architectures are unable to handle these high-frequency, large-scale data flows [11], leading to constrained data throughput and the emergence of bottlenecks [12, 13]. This limitation adversely impacts simulation efficiency and response time, ultimately degrading overall computational performance. Moreover, distributed solutions often introduce additional latency and complexity in data transmission and synchronization, further impairing simulator performance. Although NVIDIA has developed a commercial solution leveraging multi-GPU interconnection n (based on NVLink and NVSwitch), this approach remains prohibitively expensive [14] and does not fundamentally resolve the issue [15]. Significant East-West data traffic persists, offering only marginal improvements in the scalability of quantum simulators. Consequently, no effective solutions currently exist for further scaling quantum simulators.

In conclusion, the development of a high-capacity, cost-efficient, and high-throughput distributed architecture tailored for quantum simulators has emerged as a pressing necessity.

To address the constraints of existing quantum simulation architectures, this paper introduces a novel Distributed Shared Layered Storage Quantum Simulator (DSLSQS). DSLSQS employs a cooperative framework wherein multiple computational nodes share a common disk pool, effectively mitigating network transmission bottlenecks prevalent in conventional distributed systems. This architecture facilitates the efficient simulation of large-scale quantum systems while substantially reducing computational costs. Specifically, DSLSQS capitalizes on its distributed shared architecture to allow computational nodes to share a unified data space, thereby eliminating East-West traffic among nodes and alleviating bottlenecks in distributed quantum simulation computations. Additionally, DSLSQS integrates layered storage technology into its distributed shared framework, ensuring the efficient operation of critical system components while optimizing cost efficiency. Furthermore, DSLSQS eliminates TCP/IP networking overhead, thereby reducing network latency, enhancing bandwidth utilization, and further improving data transmission efficiency within the distributed architecture.

The Distributed Shared Layered Storage Quantum Simulator (DSLSQS) introduced in this paper substantially improves the simulation efficiency of large-scale quantum systems while reducing computational costs. This advancement offers essential support for investigating complex quantum phenomena and verifying quantum algorithms. The main contributions are as follows:

1) This paper conducts a comprehensive theoretical and quantitative analysis of the critical bottlenecks in high-qubit quantum state simulation, with a particular focus on the disparity between data throughput and computational power. Through rigorous quantitative analysis, we demonstrate that, with high-performance computing nodes, data transmission throughput within the distributed architecture constitutes a fundamental performance bottleneck. This insight establishes a theoretical foundation for optimizing distributed quantum simulation system.

2) This paper proposes a novel distributed shared quantum simulation system to address the inherent limitations of existing distributed architectures in quantum simulator applications. The system adopts distributed shared storage architecture within the quantum simulator [16-18]. A backend shared distributed storage pool based on CXL/PCIe/SAS network, enables unified management and coordination of resources across multiple storage devices. Furthermore, all servers have direct access to each disk. This architecture facilitates seamless data transmission across multiple computational nodes, ensuring efficient handling of large-scale data exchanges. By eliminating the East-West traffic bottleneck in distributed systems, it significantly enhances overall performance and efficiency.

3) This paper introduces a layered storage approach to reduce the cost of quantum simulators while structuring the data I/O process into a pipeline. By maintaining consistent data throughput across pipeline stages, this approach minimizes costs while preserving system performance. When integrated with the distributed shared architecture, this method theoretically reduces or even eliminates the dependency on costly DRAM/HBM, enabling the use of low-cost storage media for managing the large-scale data storage demands of quantum simulations This approach effectively mitigates the challenges posed by the exponential growth of data space, substantially reducing overall system costs while ensuring that performance degradation remains within an acceptable range.

4) This paper presents a De-TCP/IP networking approach and leverages CXL/PCIe/SAS network to optimize data transmission performance during large-scale data traversal, thereby fulfilling the rigorous data transfer requirements of quantum simulation systems.

5) Through 27-qubit quantum experiment confirms the substantial advantages of DSLSQS in practical applications. Compared to conventional quantum simulation approaches, the proposed system introduces breakthrough improvements in data I/O optimization and computational efficiency. Compared to conventional quantum simulation approaches, the proposed system introduces breakthrough improvements in data I/O optimization and computational efficiency.

The structure of this paper is as follows: Section 2 demonstrates the performance and cost bottlenecks of the distributed quantum simulator cluster. Section 3 provides a detailed discussion of the research content. Section 4 presents the experimental testing and analysis. Section 5 summarizes the research conclusions and potential future directions.

II. CURRENT RESEARCH BOTTLENECK THEORY ANALYSIS

In distributed quantum simulation research, as the number of qubits being simulated increases, the resources of a single server gradually become insufficient to meet the demands. For full-amplitude quantum simulators, when the number of qubits reaches approximately 35, it exceeds the capabilities of a single server. Although tensor quantum simulators and partial-amplitude quantum simulators can significantly reduce the demand for data space, the same issue arises when the number of simulated qubits increases and the required data space exceeds the resources of a single machine. Therefore, utilizing a distributed system composed of multiple servers to expand quantum simulation capabilities has become a key technological challenge.

In this section, the limitations of the existing system will be analyzed in detail from the aspects of technical bottlenecks and costs, and the root causes of these bottlenecks will be clearly revealed through specific mathematical models and practical cases.

*A. Technical bottleneck analysis*

1) **Computing power analysis**

To evaluate the computational capabilities of different hardware in distributed quantum simulators, we analyzed the quantum gate operation capabilities of GPU, CPU, and FPGA. Assuming the number of qubits in the system is n = 40, which simulates more complex quantum states (with the single machine's bottleneck around 34 qubits), the corresponding quantum state representation requires $2^{40}$ complex numbers. Each complex number consists of a real part and an imaginary part, each occupying 8 bytes, totaling 16 bytes.

For a simple single-qubit operation (e.g., the Hadamard gate), we need to perform addition and multiplication operations on each complex number. Each operation involves two additions and two multiplications, totaling four floating-point operations. Based on this, the total number of floating-point operations is $4 \times 2^n$, which, for n = 40, results in $4 \times 2^{40}$ floating-point operations.

GPU Computational Power: High-performance GPUs can achieve computational power on the order of $10^{15}$ FLOPS, indicating their strong parallel computing capabilities, enabling the completion of 227 quantum gate operations per second.

CPU Computational Power: The peak computational power of modern high-end server CPUs is relatively limited compared to GPUs. For example, the Intel Xeon Gold 6348 processor achieves up to 4.6 TFLOPS (single-precision) and 2.3 TFLOPS (double-precision) with its 28 cores operating at 2.6 GHz under AVX-512 support. Modern server CPUs can complete 1.15 quantum gate operations per second.

FPGA Computational Power: High-performance FPGAs have computational power on the order of $10^{12}$ FLOPS, typically suited for hardware acceleration of specific tasks. However, their overall computational capability is still lower than that of high-performance GPUs.

From the comparison above, it is evident that GPUs have significantly higher computational power than CPUs and FPGAs, making them the most efficient for quantum simulation. Specifically, GPUs can effectively accelerate quantum gate operations and meet the requirements for large-scale quantum state computations, while CPUs and FPGAs exhibit more noticeable performance bottlenecks when handling complex quantum states.

2) **Data throughput analysis**

In distributed quantum simulation, the I/O of quantum state data remains a critical bottleneck. For 40 qubits, the system represents $2^{40}$ complex numbers. Because each complex number is 16 bytes, the total data volume reaches $2^{40} \times 16 =$ 17,592,186,044,416 bytes (about 17,592 GB). Under mainstream distributed networks, a typical data transmission rate of 10 Gb/s yields a maximum throughput of about 1.25 GB/s. Even if the network speed is increased tenfold to 12.5 GB/s, transmitting the entire quantum state would still require about 1,407 seconds.

Compared with a GPU that can handle 227 quantum gates per second, the network can only transmit about 1/1407.36 ($\approx$ 0.00071) quantum gate data per second. This gap amounts to at least an order-of-magnitude difference, even relative to the weakest CPU's computational power. Consequently, data throughput emerges as the primary bottleneck, substantially restricting the quantum simulator's overall performance.

A comparative analysis of computational power and data throughput shows that computational power is not the main bottleneck in distributed quantum simulators. While GPUs offer robust computational capabilities, data throughput in conventional distributed networks (e.g., 10 Gb/s) remains severely constrained, thereby limiting the simulator's overall performance. Even increasing the network speed to 100 Gb/s fails to bridge the gap in data throughput relative to computational power.

*B. Cost analysis*

The use of quantum simulation in artificial intelligence (AI) training has received increasing attention, especially for complex model development. To thoroughly evaluate the feasibility of existing quantum simulators for AI model training, we developed a parameterized model that quantifies training iterations, computational unit configurations, and quantum gate operation frequencies, and estimates their associated storage and computational resource usage.

1) **Parameter setting during AI training**

In AI training tasks that utilize quantum simulators, we begin by identifying the essential parameters of the training process. These parameters include the number of iterations, the quantum gate operations per iteration, and the total training duration. Considering model complexity, a simple training process involves at least 10,000 iterations, with around 1,000 quantum gate operations each, to mimic forward and backward propagation in AI training. If the entire training completes within 24 hours, the total duration amounts to 86,400 seconds.

2) **Calculate the operating frequency of quantum gates**

Based on the above Settings, the total number of quantum gate operations can be expressed as:

$$Q_{total} = m_{iter} \times n_{gate}, \tag{1}$$

4where, $m_{iter}$ represents the number of iterations, and $n_{gate}$ represents the number of quantum gate operations per iteration.

In order to meet the real-time training needs, the quantum gate operating frequency per second needs to be calculated:

$$Q_{per} = \frac{Q_{total}}{t}, \quad (2)$$

where, $t$ represents the training time, $Q_{per}$ represents the number of quanta the system needs to process per second.

In the case of a 40-qubit simulation, this frequency indicates that the system needs to process about 116 quantum-gate operations per second.

### 3) Computing resource requirements: GPU and CPU configuration

Quantum simulation entails a substantial volume of floating-point operations, particularly for 40-qubit gate operations. A single quantum gate on 40 qubits requires roughly $4 \times 2^{40}$ FLOPS. For instance, a high-end GPU like the NVIDIA A100 or H100 delivers around $10^{15}$ floating-point operations per second, enabling it to process the following number of quantum gates per second:

$$G_{GPU} = \frac{10^{15}}{4 \times 2^{40}} \approx 227.37. \quad (3)$$

In comparison, a high-end CPU like the Intel Xeon Gold 6348 achieves up to 4.6 TFLOPS (single-precision) or 2.3 TFLOPS (double-precision) under AVX-512. Given that a single quantum gate on 40 qubits requires approximately $4 \times 2^{40}$ FLOPS (single-precision), the number of quantum gates the CPU can process per second is:

$$G_{CPU} = \frac{4.6 \times 2^{40}}{4 \times 2^{40}} = 1.15. \quad (4)$$

This results in approximately 1.15 quantum gates per second, significantly lower than high-performance GPUs, highlighting the performance gap for large-scale quantum simulations.

Because the system needs to handle 116 quantum gate operations per second, a single GPU's computational power is already adequate to satisfy this requirement. In contrast, a high-end CPU would struggle to achieve the same performance. Hence, deploying a single GPU is enough to achieve the necessary computational performance.

### 4) Storage requirements and cost analysis

For a 40-qubit system, the quantum state data comprises $2^{40}$ complex numbers, each requiring 16 bytes, leading to about 16 TB of total storage. To address this need, our study analyzes the storage costs of existing solutions:

- DRAM Storage: High-performance DRAM (e.g., DDR4/DDR5) currently costs about 5 USD/GB, totaling roughly 82,000 USD.
- HBM Storage: GPUs may employ high-bandwidth memory (HBM), priced at about 20 USD/GB, for an overall cost of nearly 330,000 USD.
- CPU Memory (DDR5 ECC RAM): High-performance server CPUs rely on DDR5 ECC RAM, which costs approximately 6 USD/GB, leading to a total cost of 98,000 USD.

### 5) Cost structure and analysis

Given the aforementioned computational resources and storage needs, a GPU configuration costs around 10,000 USD, whereas DRAM storage alone reaches about 82,000 USD—nearly ten times higher. If HBM storage is employed, the storage cost soars to 33 times that of the GPU configuration.

For CPU-based simulations, the computational cost of a high-performance CPU server is approximately 15,000 USD. However, the memory cost (DDR5 ECC RAM) almost 100,000 USD. This highlights that storage costs remain the dominant expense in both CPU and GPU configurations.

Evidently, storage costs comprise the dominant share of the quantum simulator's total expense. Reducing the overall cost of quantum simulators thus requires strategies to curb storage expenses for the vast data space demanded by quantum simulation.

This section examines the storage costs of quantum simulators in AI training, which substantially exceed those of computational units. In light of low-latency and high-throughput requirements, HBM represents the best option to meet performance needs. However, because HBM remains substantially more expensive than GPUs, future quantum simulator architectures must urgently seek solutions that combine efficient storage and cost-effectiveness to reduce expenses while still meeting computational requirements.

## III. OUR PROPOSED METHOD

In response to the data throughput bottleneck in quantum simulation, we propose a novel Distributed Shared Layered Storage Quantum Simulator, named DSLSQS. In this section, we provide a detailed overview of DSLSQS's design philosophy, architectural features, and core technologies.

### A. Total framework

The distributed shared quantum simulator infrastructure proposed here is structured into an application layer, a software layer, and an underlying hardware layer. The overall architecture design is shown in Figure 1:

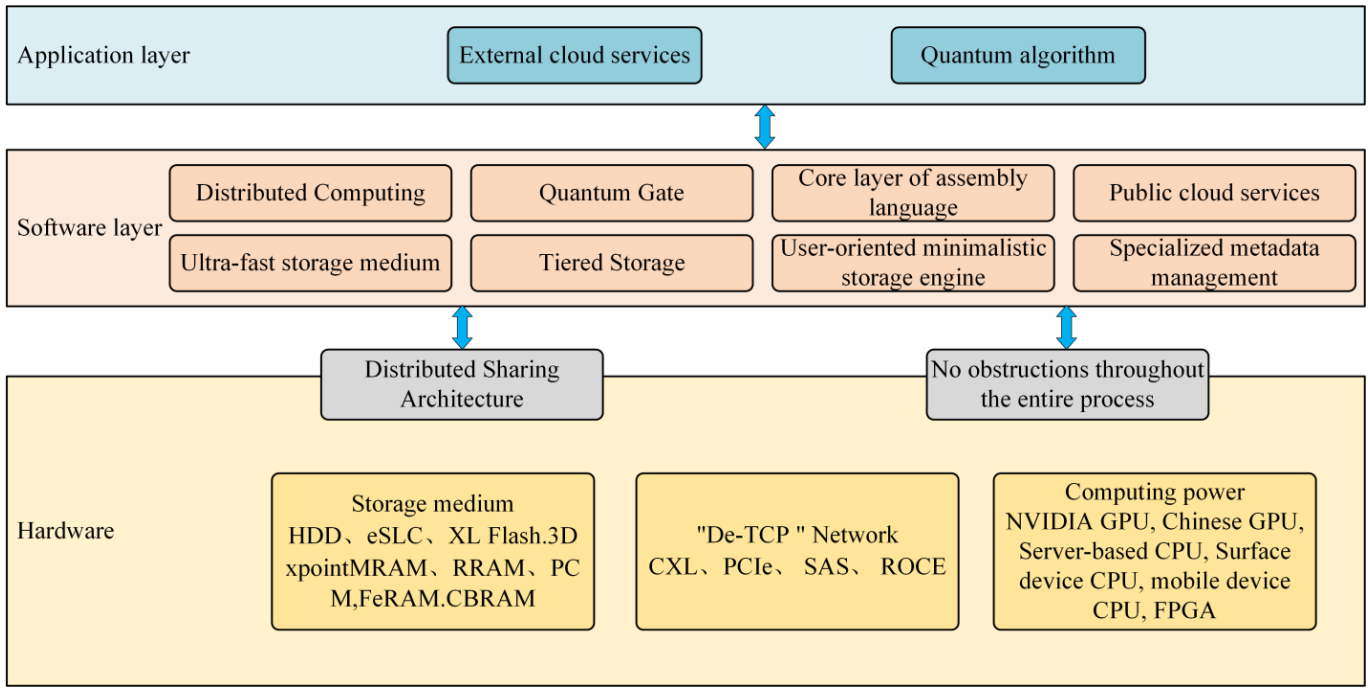

**Fig. 1.** Overall architecture of distributed shared quantum simulator

The application layer primarily targets users and developers, serving as the main interface with the quantum simulator. This layer offers an extensive suite of tools and interfaces, enabling users to debug and execute quantum algorithms and to access external cloud services. Additionally, this layer incorporates various optimization libraries and algorithm frameworks, enabling quantum computing researchers and engineers to develop applications efficiently without dealing directly with the underlying architecture.

The core software layer oversees the quantum simulation process, encompassing quantum gate computation, distributed task scheduling, metadata management, storage engines, distributed data coordination, layered storage, compilation, and optimization.

The underlying hardware layer functions as the physical core of the architecture, delivering computational, storage, and network resources essential for quantum simulation. This layer integrates high-performance classical computing resources—CPUs, GPUs, and FPGAs—to enable large-scale parallel computation. It also incorporates storage media for quantum simulation data and network devices that link these storage resources with computational units. The underlying hardware must ensure computational efficiency while meeting the complex storage and transmission demands inherent in quantum simulation. To accommodate large-scale quantum states, the hardware implements both a distributed shared storage architecture and a layered storage scheme, guaranteeing real-time data transmission and sustained computational speed throughout the simulation.

*B. Distributed shared architecture*

Because quantum gate operations in distributed quantum simulator cluster are highly complex, node-to-node communication is especially critical. The system's network topology governs both the physical and logical relationships among nodes, as well as the methods and efficiency of quantum circuit simulation. Existing technologies generally load data into specific server memory and then transmit it to other servers over TCP/IP networks. As shown in Figure 2, this paper presents a distributed shared architecture in which every computational node has access to the entire dataset directly, without help of another node.

Consequently, the system experiences only North-South data traffic, while East-West traffic is confined to minimal signaling that does not affect performance. Under this architectural framework, instead of residing on computational servers, the storage medium is directly connected to the network. All computational servers can then access it directly, forming a back-end shared distributed storage pool that supports efficient data management and retrieval. In this architecture, data flow is optimized, substantially shortening the path from storage media to computational units and attaining performance close to local disk read/write speeds. Crucially, data no longer must travel between computational nodes, thus removing East-West data traffic in the distributed system and breaking through the bottleneck in distributed quantum simulation.

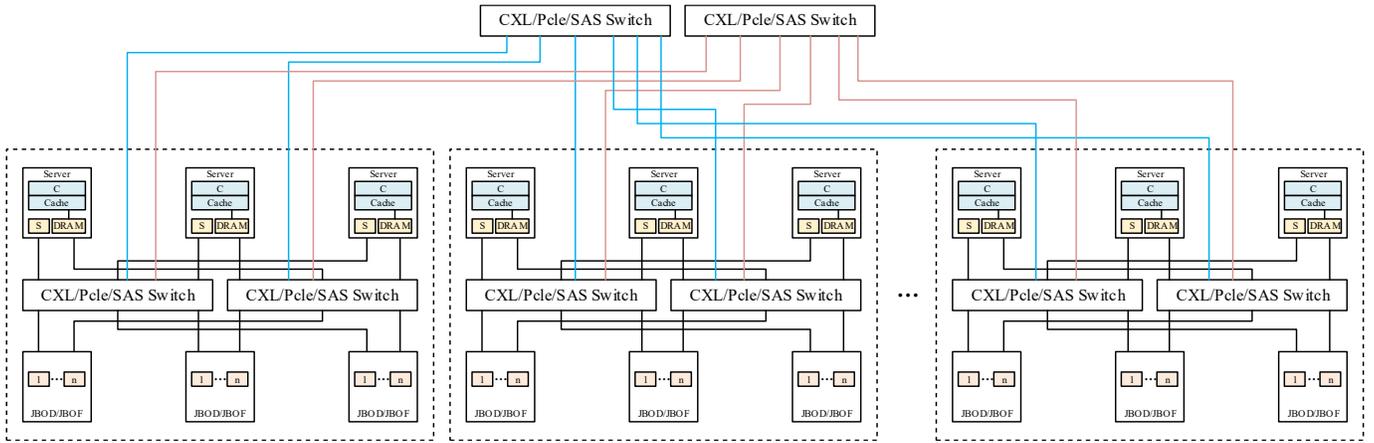

**Fig. 2.** System link topology

*C. Hardware architecture*

Hardware architecture addresses the physical makeup of a computer system and the design and organization of its internal components. It encompasses the functionality of hardware components—like processors, memory, storage devices, and I/O devices—alongside their design and operating mechanisms. An efficient hardware architecture ultimately dictates the system's data-processing efficiency. By employing a distributed shared architecture, any computational unit (GPU, CPU etc.) can directly transfer data to or from any storage unit (memory, disk, etc.), thus enabling fast computation and efficient data access.

The distributed shared quantum simulator cluster seamlessly operates through the coordinated interplay of storage, computing, and communication devices, non-blocking throughout the entire process, fully harnessing its synchronization capabilities. As illustrated in Figure 3, the architecture integrates multiple storage media with varied performance tiers. However, with careful design, the system forms a data I/O pipeline in which each stage may exhibit substantially different latencies, but overall throughput remains nearly constant, resulting in a non-blocking I/O path.

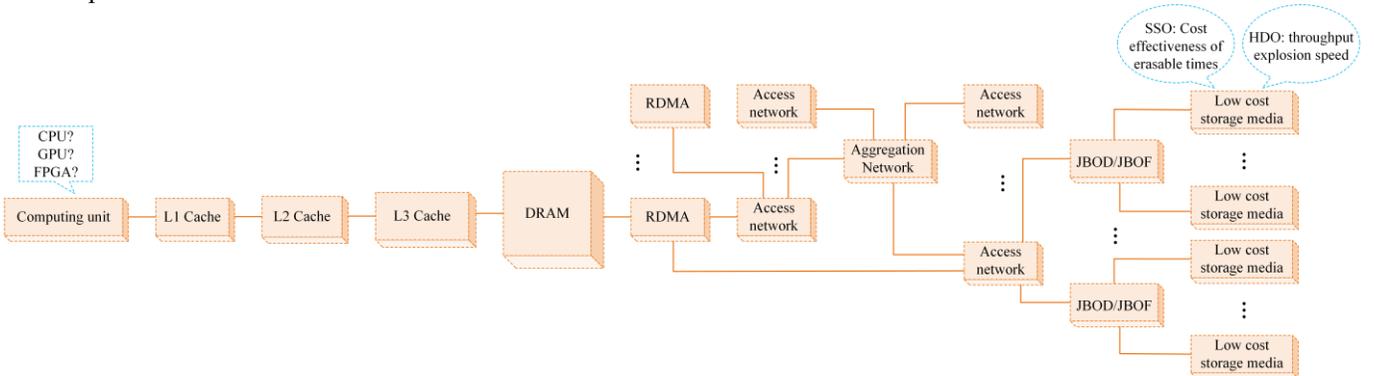

**Fig. 3.** Hardware architecture

*D. Key technology*

**1) Distributed shared storage technology**

As illustrated in Figure 4, the distributed shared storage approach directly connects storage media—excluding system disk for OS & software—to the network, typically via storage protocols like SAS, PCIe, or CXL, which is usually the same interface as the storage media. This storage network links each computational node to a back-end shared distributed storage pool, enabling seamless collaboration among distributed nodes and resulting in a highly optimized data storage system tailored to quantum simulators.

This system deeply optimizes the data transmission path, effectively reducing the frequent data interactions and traffic overhead between servers in traditional distributed systems. Unlike existing architectures, in which computational nodes depend on one another for data access, this system decouples computational nodes from the storage layer. The storage media is not tied to any single node but is shared across all nodes. Any computational node can directly read from or write to any storage medium, eliminating the need to route data through other nodes.

The key to this design is that each server can bypass other servers and directly access all data in the storage pool. This architecture removes the horizontal East-West data traffic inherent in conventional distributed systems, thereby ensuring stable and efficient data access. This direct-access model substantially boosts overall system throughput, fulfilling the high-throughput demands of distributed quantum simulators in large-scale computations.

By implementing the aforementioned optimizations, the distributed shared quantum simulation cluster delivers enhanced performance, offering robust support for stable and efficient large-scale quantum simulation tasks.

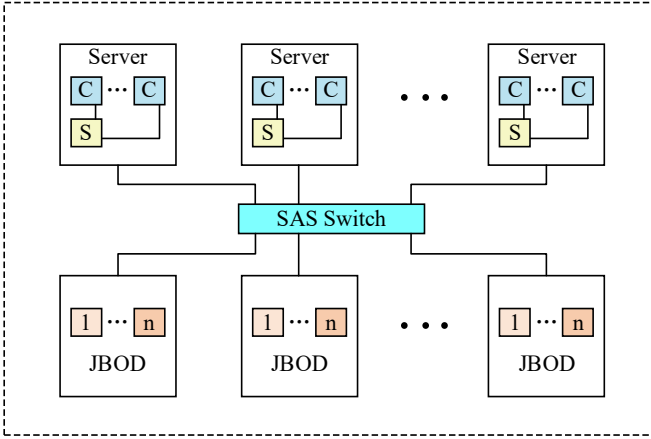

**Fig. 4.** Distributed shared storage technology

**2) Hierarchical storage technology**

Traditional hierarchical storage strategies are not fully suitable for quantum simulation systems due to the necessity of quantum data traversal. In quantum simulator applications, computational resources have ceased to be the primary bottleneck. Instead, data throughput has emerged as the critical factor limiting overall system efficiency. Efficient data transfer capabilities are thus essential for maintaining quantum simulation performance. Nevertheless, existing quantum simulator architectures incur significant costs in addressing these demands.

To balance performance and cost, we propose a novel layered storage approach integrated with distributed shared storage technology. Computational units are deployed in high-performance storage regions, ensuring that data throughout matches that of computational units. The permanent storage of data is allocated to cost-effective storage areas, optimizing the cost-efficiency and resource utilization of the system. During computation, data is transferred from permanent storage media directly into the cache of computational units, theoretically reducing or even removing dependence on expensive DRAM or HBM.

Throughout the computation process, the system maintains consistent throughput at each data transmission stage. Both the scheduling of data from high-performance storage to computational units and the data flow from cost-effective storage areas to the storage network have been rigorously optimized and coordinated. This design effectively eliminates bottlenecks and imbalances within the data transmission path, thereby preventing performance degradation caused by mismatch of different transmission stages.

The storage system integrates storage networks and devices, fully leveraging storage media characterized by high rewrite durability and moderate cost. This integration maximizes throughput and ensures efficient data storage and retrieval. Under this architecture, the system not only addresses large-scale data access requirements but also optimizes data transmission paths, maintaining network throughout and costs within acceptable limits. The high-performance storage regions guarantee real-time execution of critical computational tasks, whereas the incorporation of cost-effective storage regions significantly reduces overall operational costs. By balancing performance and cost-efficiency, this solution substantially improves overall computational efficiency and resource utilization, strongly supporting high-load quantum simulation tasks.

**3) De-TCP /IP networking technology**

Existing distributed systems typically rely on TCP/IP protocols, originally designed decades ago for dial-up telephone networks. These protocols are inherently complex and inefficient, exhibiting typical latency of approximately 1 ms.

De-TCP /IP networks, including SAS, PCIe, CXL, InfiniBand, and RoCE networks, have performance generally below 10μs, and some can even reach below 1μs. This represents a latency improvement ranging from tens to thousands of times over conventional distributed systems.

Furthermore, SAS, PCIe, and CXL (based on PCIe) inherently serve as storage interfaces, allowing storage media to directly connect to the network without protocol conversion, thus enabling even greater performance.

This paper employs a De-TCP/IP network for storage connectivity, fully harnessing the superior capabilities of the distributed storage architecture.

## IV. EXPERIMENTAL TEST AND COMPREHENSIVE ANALYSIS

*A. Experimental environment configuration and hardware parameter setting*

This experiment was conducted in a distributed computing environment to compare the performance of existing distributed systems and the proposed distributed shared storage architecture in supporting quantum simulator operations. The experiment included an experimental group and a control group, each consisting of two computational nodes with identical configurations, as detailed below:

- Processor: Intel Xeon E5-2660v3 (4 cores);
- Memory: 8GB;
- Storage: SSD;
- Network: Gigabit Ethernet;

This experiment employs specially designed quantum simulation software with layered storage support. The software stores data within a distributed storage system instead of memory, markedly increasing storage capacity and substantially lowering overall system costs.

In the experimental group, the proposed DSLSQS was implemented, with two nodes sharing SSD storage located in JBOD. The control group utilized the widely-used CEPH distributed storage system, operating in single-replica mode. To comprehensively assess the impact of cache configurations on performance, experiments were conducted under two cache sizes: 512 MB and 1024 MB.

*B. Experimental principle and process*

The primary goal of this experiment is to assess the performance of the proposed distributed architecture and compare it against existing distributed solutions. The experiment simulates a 27-qubit quantum system, employing a distributed quantum simulator to perform quantum gate operations across multiple qubits. The quantum gate operations performed include the Hadamard (H) and X gates.

1) **Initial status and system configuration**

   During system initialization, the 27-qubit system is initialized to the state vector |00000000000001000 000000000⟩, represented as complex numbers. The system's required state space is $2^{27}$ (134,217,728 states). Since each state is represented by a 16-byte complex number, the total storage required is approximately 2 GB. The initial state is configured on the master node (Node 0) and stored on disk storage devices.

2) **Distributed computation**

   This experiment employs a distributed computing approach, utilizing two computational nodes (Node 0 and Node 1) to process different state ranges, with Node 0 designated as the master node. Specifically, each node is responsible for computing the quantum state indices assigned to it. For example, Node 0 processes state data for indices ranging from 0 to 67,108,863, while Node 1 handles state data from 67,108,864 to 134,217,727.

3) **Communication and synchronization**

   Throughout the experiment, nodes synchronize and exchange data in real time via the network. When a node completes its quantum state computation, it reports its status to the master node through the network and waits for subsequent instructions. This distributed computing mechanism allows computational tasks to be efficiently parallelized even for high-dimensional quantum systems, thus substantially reducing overall computation time.

*C. Experimental result*

This experiment simulates a 27-qubit quantum system by performing H and X gate operations on a distributed platform, and gathers experimental data to assess system performance.

In this experiment, the two computational nodes handle computations for distinct segments of the quantum state space. While processing their respective assigned state ranges, each node simulates a 27-qubit system by sequentially applying the H and X gates to the 13th qubit. The experiment records several performance metrics, including computation time, disk read/write durations, and disk write-back latency. Tables 1 and 2 summarize the performance of each node during H gate and X gate operations.

TABLE I
PERFORMANCE OF 512M CACHE H GATE OPERATION AND X GATE OPERATION.

| Framework | Node | Quantum gate | Computing time (ms) | Disk read time (ms) | Disk write time (ms) | Writing the disk back time (ms) | Total time (ms) | Computation speed (mb/s) |
|---|---|---|---|---|---|---|---|---|
| DSLSQS | Node 0 | Quantum gate 1 (H gate) | 811 | 4109 | 466 | 2281 | 7671 | 133 |
|  |  | Round-0 | 9575 |  |  |  |  | 213 |
|  |  | Quantum gate 2 (X gate) | 787 | 4626 | 444 | 3478 | 9338 | 109 |
|  |  | Round-1 | 9968 |  |  |  |  | 205 |
|  | Node 1 | Quantum gate 1 (H gate) | 1073 | 4727 | 444 | 2532 | 8779 | 116 |
|  |  | Quantum gate 2 (X gate) | 1075 | 4634 | 447 | 3808 | 9966 | 102 |
| CEPH | Node 0 | Quantum gate 1 (H gate) | 798 | 33474 | 446 | 5334 | 40055 | 25 |
|  |  | Round-0 | 44881 |  |  |  |  | 45 |
|  |  | Quantum gate 2 (X gate) | 803 | 24041 | 446 | 9337 | 34630 | 29 |
|  |  | Round-1 | 45004 |  |  |  |  | 45 |
|  | Node 1 | Quantum gate 1 (H gate) | 1072 | 34598 | 445 | 8382 | 44500 | 23 |
|  |  | Quantum gate 2 (X gate) | 1073 | 38650 | 448 | 4829 | 45003 | 22 |

TABLE II
PERFORMANCE OF 1024M CACHE H GATE OPERATION AND X GATE OPERATION.

| Framework | Node | Quantum gate | Computing time (ms) | Disk read time (ms) | Disk write time (ms) | Writing the disk back time (ms) | Total time (ms) | Computation speed (mb/s) |
|---|---|---|---|---|---|---|---|---|
| DSLSQS | Node 0 | Quantum gate 1 (H gate) | 739 | 4722 | 392 | 3217 | 9071 | 112 |
|  |  | Round-0 | 9999 |  |  |  |  | 204 |
|  |  | Quantum gate 2 (X gate) | 664 | 4676 | 360 | 3279 | 8980 | 114 |
|  |  | Round-1 | 9805 |  |  |  |  | 208 |
|  | Node 1 | Quantum gate 1 (H gate) | 1059 | 4593 | 444 | 3525 | 9623 | 106 |
|  |  | Quantum gate 2 (X gate) | 1066 | 4735 | 445 | 3557 | 9803 | 104 |
| CEPH | Node 0 | Quantum gate 1 (H gate) | 754 | 22978 | 372 | 21343 | 45449 | 22 |
|  |  | Round-0 | 45449 |  |  |  |  | 45 |
|  |  | Quantum gate 2 (X gate) | 778 | 22834 | 377 | 17484 | 41475 | 24 |
|  |  | Round-1 | 45449 |  |  |  |  | 45 |
|  | Node 1 | Quantum gate 1 (H gate) | 1065 | 22970 | 449 | 16469 | 40955 | 25 |
|  |  | Quantum gate 2 (X gate) | 1066 | 22023 | 449 | 23845 | 47383 | 21 |

*D. Experimental analysis and discussion*

**1) Benefits of DSLSQS:**

The proposed distributed shared storage architecture significantly outperforms CEPH, a state-of-the-art distributed storage technology. DSLSQS achieves an average throughput of 207.5 MB/s, compared to 44.5 MB/s for CEPH — an improvement of approximately 366%. I Regarding disk read/write performance, DSLSQS exhibits substantially lower latency, averaging 4602.75 ms for disk reads and 3209.625 ms for disk write-back operations. In contrast, CEPH shows significantly higher latency, with an average disk read time of 27,696 ms and write-back latency averaging 13,377.875 ms. These results clearly demonstrate that the proposed architecture is highly suitable for high-performance, data-intensive computational tasks, including quantum simulation.

The experimental results demonstrate that the DSLSQS architecture effectively overcomes the performance bottlenecks of distributed systems, thereby enhancing computational efficiency.

**2) System scalability and availability:**

The experiment results clearly demonstrate that the distributed shared storage architecture possesses excellent scalability and usability for large-scale quantum simulations. As the number of qubits increases, the proposed architecture can provide enough computing resources and work together in a multi-node environment to maintain high computational efficiency. In contrast, traditional distributed systems demonstrate suboptimal performance in large-scale parallel computations, potentially failing to satisfy the requirements of future quantum computing applications.

**3) Impact of cache configuration on computing performance:**

When employing a larger cache (1024 MB), neither storage architecture showed a noticeable improvement in computational speed. While increasing cache size typically enhances access performance for frequently accessed data, the traversal characteristic of quantum gate computations results in uniform access patterns throughout the dataset. Consequently, no frequently accessed ("hot") data regions can be identified, making caching largely ineffective.

In summary, this experiment validates the advantages of the DSLSQS architecture in quantum simulation, clearly demonstrating its capability to markedly improve computational speed and substantially reduce data I/O latency through the adoption of layered storage. Consequently, the DSLSQS architecture achieves superior performance and scalability, effectively handling complex quantum computing tasks. In contrast, existing distributed systems encounter significant performance bottlenecks, resulting in lower computational efficiency and notably inferior performance relative to the DSLSQS approach.

It is important to note that the quantum simulation software supporting layered storage employed in this experiment remains at the prototype stage, with performance optimizations still underway. Therefore, the absolute performance metrics presented here have not yet reached their optimal levels. Nevertheless, these results are sufficient to demonstrate the feasibility of layered storage and clearly highlight the distinct advantages of the distributed shared storage architecture compared with traditional distributed storage systems.

V. CONCLUSION AND PROSPECT

This paper examines technical bottlenecks in existing distributed quantum simulators for large-scale quantum system simulations and proposes a novel architecture, the Distributed Shared Layered Storage Quantum Simulator (DSLSQS). In conventional quantum simulator architectures, scalability regarding the number of qubits is predominantly restricted by data I/O, making it the primary performance bottleneck of the system. Quantitative analysis conducted in this study demonstrates that overall system efficiency is primarily limited by data throughput rather than computational power, highlighting storage and data transmission constraints as more critical factors.

To overcome these limitations, this study proposes an innovative solution comprising the following core components: First, a novel distributed shared storage architecture is proposed, enabling multiple computational nodes to directly access a unified storage pool. This approach eliminates the East-West data traffic bottlenecks present in conventional distributed systems, substantially improving data access efficiency and addressing scalability constraints in quantum simulation computations. Second, a layered storage technology is employed to further optimize cost efficiency. While maintaining high performance, this method utilizes cost-effective storage media to manage large-scale quantum state data, substantially reducing dependence on expensive memory solutions like DRAM and HBM. This innovative framework effectively addresses the exponential growth in data storage requirements for quantum simulations, achieving substantial cost reductions without sacrificing system performance. Furthermore, this study adopts a TCP/IP-free networking paradigm, integrating high-speed communication protocols including CXL, PCIe, and SAS to optimize data transmission pathways, thus maximizing throughput and minimizing network latency. By integrating these technologies, DSLSQS significantly enhances the capabilities of distributed quantum computing architectures, providing a scalable and cost-effective solution for large-scale quantum simulations.

Through the simulation of a 27-qubit experiment, the proposed distributed shared storage quantum simulation system demonstrated significant advantages in practical applications. Experimental results from the 27-qubit quantum system simulation indicate that the DSLSQS architecture effectively reduces data I/O latency and substantially enhances overall computational speed, achieving a performance improvement of approximately 366%.

In conclusion, this paper introduces a novel architecture that integrates distributed shared storage with layered storage, offering an efficient and cost-effective solution for the scalable deployment of quantum simulators. This work establishes a strong foundation for advancing quantum computing research and practical applications. Future research could further enhance the DSLSQS architecture by maximizing data I/O performance and investigating additional strategies tailored for large-scale quantum simulations.

REFERENCES

[1] L. Gyongyosi and S. Imre, "A Survey on quantum computing technology," *Computer Science Review,* vol. 31, pp. 51-71, Feb 2019, doi: 10.1016/j.cosrev.2018.11.002.
[2] M. Horowitz and E. Grumbling, "Quantum computing: progress and prospects," 2019.
[3] B. Fauseweh, "Quantum many-body simulations on digital quantum computers: State-of-the-art and future challenges," *Nature Communications,* vol. 15, no. 1, Mar 2024, Art no. 2123, doi: 10.1038/s41467-024-46402-9.
[4] J. Singh and K. S. Bhangu, "Contemporary Quantum Computing Use Cases: Taxonomy, Review and Challenges," *Archives of Computational Methods in Engineering,* vol. 30, no. 1, pp. 615-638, Jan 2023, doi: 10.1007/s11831-022-09809-5.
[5] B. P. Lanyon *et al.*, "Universal digital quantum simulation with trapped ions," *Science,* vol. 334, no. 6052, pp. 57-61, 2011.
[6] A. J. Daley *et al.*, "Practical quantum advantage in quantum simulation," *Nature,* vol. 607, no. 7920, pp. 667-676, 2022.
[7] M. C. Tran, Y. Su, D. Carney, and J. M. Taylor, "Faster digital quantum simulation by symmetry protection," *PRX Quantum,* vol. 2, no. 1, p. 010323, 2021.
[8] F. Tacchino, A. Chiesa, S. Carretta, and D. Gerace, "Quantum Computers as Universal Quantum Simulators: State-of-the-Art and Perspectives," *Advanced Quantum Technologies,* vol. 3, no. 3, Mar 2020, Art no. 1900052, doi: 10.1002/qute.201900052.
[9] A. Broadbent, J. Fitzsimons, and E. Kashefi, "Universal blind quantum computation," in *2009 50th annual IEEE symposium on foundations of computer science*, 2009: IEEE, pp. 517-526.
[10] D. Cuomo, M. Caleffi, and A. S. Cacciapuoti, "Towards a distributed quantum computing ecosystem," *IET Quantum Communication,* vol. 1, no. 1, pp. 3-8, 2020.
[11] X. P. Zhou, Z. L. Liu, M. Z. Guo, J. C. Zhao, and J. L. Wang, "SACC: A Size Adaptive Content Caching Algorithm in Fog/Edge Computing Using Deep Reinforcement Learning," *IEEE Transactions on Emerging Topics in Computing,* vol. 10, no. 4, pp. 1810-1820, Oct 2022, doi: 10.1109/tetc.2021.3115793.
[12] S. Ghanavati, J. Abawajy, and D. Izadi, "Automata-Based Dynamic Fault Tolerant Task Scheduling Approach in Fog Computing," *IEEE Transactions on Emerging Topics in Computing,* vol. 10, no. 1, pp. 488-499, Jan 2022, doi: 10.1109/tetc.2020.3033672.
[13] S. D. di Vimercati, S. Foresti, S. Jajodia, S. Paraboschi, P. Samarati, and R. Sassi, "Sentinels and Twins: Effective Integrity Assessment for Distributed Computation," *IEEE Transactions on Parallel and Distributed Systems,* vol. 34, no. 1, pp. 108-122, Jan 2023, doi: 10.1109/tpds.2022.3215863.
[14] M. Manathunga, H. M. Aktulga, A. W. Götz, and K. M. Merz Jr, "Quantum mechanics/molecular mechanics simulations on NVIDIA and AMD graphics processing units," *Journal of Chemical Information and Modeling,* vol. 63, no. 3, pp. 711-717, 2023.
[15] Z. W. Wang, "Blockchain-Based Edge Computing Data Storage Protocol Under Simplified Group Signature," *IEEE Transactions on Emerging Topics in Computing,* vol. 10, no. 2, pp. 1009-1019, Apr-Jun 2022, doi: 10.1109/tetc.2021.3062348.
[16] Z. Yang *et al.*, "OceanBase Paetica: A Hybrid Shared-Nothing/Shared-Everything Database for Supporting Single Machine and Distributed Cluster," *Proceedings of the VLDB Endowment,* vol. 16, no. 12, pp. 3728-3740, 2023.
[17] D.Wang, Y.Jin and Y.Q, "Storage system", Patent CN105472047A Patent Appl. CN201610076422.6.
[18] Y.Jin and H.Zhou, " Method and apparatus for a virtual machine to access storage devices in a cloud computing management platform", Patent CN105808165A Patent Appl. CN201610120933.3

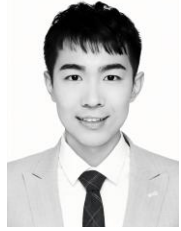

**MINGYANG YU** received his bachelor's degree from Jinan University's School of International Energy in 2022. Currently, he is pursuing a Ph.D. at the College of Artificial Intelligence, Nankai University. His research interests include the architecture design and performance optimization of quantum simulators, the construction of distributed quantum simulation platforms for large-scale simulation tasks, and the development of efficient classical simulation algorithms for specific quantum systems.

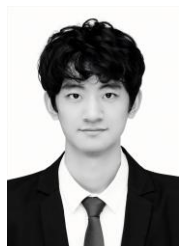

**HAORUI YANG** received the Bachelor of Engineering degree in Vehicle Engineering from Hefei University of Technology, China, in 2023. He is pursuing the Master of Engineering degree in Electronic Information at the College of Artificial Intelligence, Nankai University, China. His research focuses on quantum computing and high-performance quantum simulation, including the design and optimization of quantum simulators, distributed quantum simulation architectures, and efficient classical algorithms for simulating quantum systems.

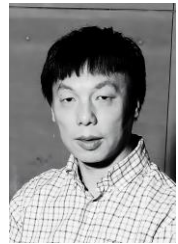

**DONGLIN WANG**, Chief Scientist at Sursen Corp. and professor at Nankai University's College of AI., has decades of engineering experience, 200+ patents in IT infrastructure, AI, blockchain, quantum computing, and digital documents, along with academic research in logic, the philosophy of science and axiomatized physics. He was Chair of the OASIS UOML-X TC, leading global standards development. He also contributed to tech legislation. He won several top national awards from CAST, MOST, and MII. Alex earned his bachelor's degree in computer science from Nankai University in 1989.

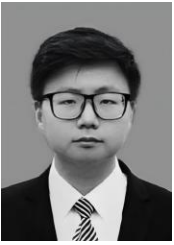
**DESHENG KONG** received his Master's degree in Software Engineering from the School of Software, Nankai University. Now he is currently pursuing his Ph.D. degree in Artificial Intelligence at the School of Artificial Intelligence, Nankai University. His primary research interests focus on quantum machine learning, quantum computing, and graph learning.

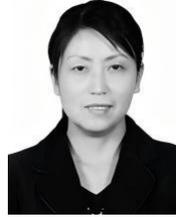
**JING XU** (Member, IEEE) is a professor at the College of Artificial Intelligence, Nankai University. She received her Ph.D. degree from Nankai University in 2003. She has published more than 100 papers in software engineering, software security, and big data analytics. She won the second prize of the Tianjin Science and Technology Progress Award twice in 2017 and 2018, respectively.

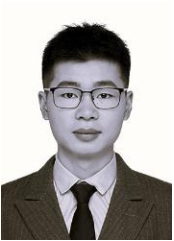
**JI DU** received his bachelor's degree from Nankai University in 2023. He is currently pursuing the Ph.D. degree with the College of Artificial Intelligence, Nankai University, Tianjin, China. His research interests include machine learning and computer vision.

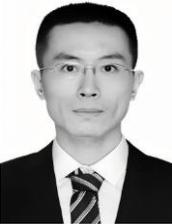
**YULONG FU** was born in June 1989. He is a senior engineer at China Electronics Technology Group Corporation and serves as Deputy Director of the Science and Technology Committee at the Yangtze Delta Region Industrial Innovation Center of Quantum and Information Technology. His research interests include quantum simulator architecture, distributed quantum simulation, and quantum software systems. He has led key projects in quantum computing networks, post-quantum cryptography, and quantum algorithms, with applications in secure communications, finance, and AI.

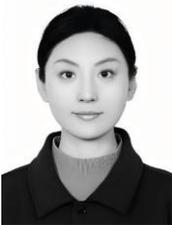
**WEI WANG**, born in March 1985, is a senior engineer specializing in quantum information system design. She serves as Special Assistant Director at the Yangtze Delta Region Industrial Innovation Center of Quantum and Information Technology and Deputy Director at the Quantum Computing Technology Development Research Center of the China Electronics Academy. Her work focuses on superconducting quantum systems, quantum software, and strategic applications. She has led key national projects, published research on quantum encryption and simulation, and holds patents in image classification and microwave interconnect technologies.